\documentclass[conference, letterpaper]{IEEEtran}
  \usepackage{caption}
  \usepackage{comment}
  \usepackage{lipsum}
\usepackage[printonlyused]{acronym}
\usepackage{adjustbox}
\usepackage[style=ieee,backend=bibtex]{biblatex}
\bibliography{biblio}
\include{macros}
\usepackage{array}

\usepackage[font=small]{caption}
\usepackage{subfigure}
\usepackage{amsmath}
\usepackage{bm}

\begin{document}

\title{Small UAVs-supported Autonomous Generation of Fine-grained 3D Indoor Radio Environmental Maps}
\author{
\IEEEauthorblockN{Ken Mendes, Filip Lemic, Jeroen Famaey}
\IEEEauthorblockA{Internet and Data Lab (IDLab), Universiteit Antwerpen - imec, Belgium} 
Email: \{name.surname\}@uantwerpen.be
\vspace{-3mm}
}

\maketitle

\begin{abstract}
\acp{REM} are a powerful tool for enhancing the performance of various communication and networked agents.
However, generating REMs is a laborious undertaking, especially in complex \ac{3D} environments, such as indoors. 
To address this issue, we propose a system for autonomous generation of fine-grained REMs of indoor 3D spaces. 
In the system, multiple small indoor \acp{UAV} are sequentially used for 3D sampling of signal quality indicators.
The collected readings are streamlined to a \ac{ML} system for its training and, once trained, the system is able to predict the signal quality at unknown 3D locations.
The system enables automated and autonomous REM generation, and can be straightforwardly deployed in new environments.
In addition, the system supports REM sampling without self-interference and is technology-agnostic, as long as the REM-sampling receivers features suitable sizes and weights to be carried by the \acp{UAV}.   
In the demonstration, we instantiate the system design using two UAVs and show its capability of visiting 72 waypoints and gathering thousands of Wi-Fi data samples. 
Our results also include an instantiation of the ML system for predicting the \ac{RSS} of known Wi-Fi \acp{AP} at locations not visited by the UAVs.
\end{abstract}



\acrodef{SDMs}{Software-Defined Metamaterials}
\acrodef{SDM}{Software-Defined Metamaterial}
\acrodef{THz}{Terahertz}
\acrodef{UWB}{Ultra Wide-Band}
\acrodef{FPGA}{Field Programmable Gate Array}
\acrodef{ToF}{Time of Flight}
\acrodef{AoA}{Angle of Arrival}
\acrodef{RSS}{Received Signal Strength}
\acrodef{AP}{Access Point}
\acrodef{3D}{3-Dimensional}
\acrodef{TS-OOK}{Time-Spread ON-OFF Keying}
\acrodef{RF}{Radio Frequency}
\acrodef{REM}{Radio Environmental Map}
\acrodef{USB}{Universal Serial Bus}
\acrodef{ISM}{Industrial, Scientific, and Medical}
\acrodef{RSS}{Received Signal Strength}
\acrodef{SUT}{System Under Test}
\acrodef{SNR}{Signal-to-Noise Ratio}
\acrodef{CSI}{Channel State Information}
\acrodef{ML}{Machine Learning}
\acrodef{UAV}{Unmanned Aerial Vehicle}
\acrodef{GPS}{Global Positioning System}
\acrodef{LPS}{Loco Positioning System}
\acrodef{RMSE}{Root Mean Square Error}
\acrodef{MAC}{Medium Access Control}
\acrodef{SSID}{Service Set Identifier}
\acrodef{LPS}{Loco Positioning System}
\acrodef{LPD}{Loco Positioning Deck}

\section{Introduction}

A \acf{REM} documents radio signal properties over a given geographic area. 
These properties can among others include the frequency, protocol and technology of the radio signal, as well as an indication of the quality of the signal (e.g., \acf{RSS}, \acf{SNR}, or \acf{CSI}), and are stored together with the location where they were measured. 
These REMs and the data they hold can then be used for a variety of purposes, for example as an aid in cognitive radio networks~\cite{yilmaz2013radio}, for RF-based localization~\cite{lemic2016enriched}, or for optimizing network discovery and handover procedures~\cite{santi2021location}.

As a trade-off to their utility, the generation of REMs is a burdensome process, particularly for \ac{3D} environments~\cite{caso2019vifi}. 
Apart from the significant labor needed for carrying out a measurement campaign, an additional complexity stems from the fact that the measurements have to be correlated with the physical locations at which they are collected, suggesting the need for accurate localization of the entity collecting such measurements. 
The complexity of REM generation is further exacerbated by the fact that such generation often has to be performed periodically, as the REMs can become obsolete due to long-term changes in the signal propagation~\cite{meshkova2011experimental}.
Finally, as a variety of different wireless technologies often co-exist~\cite{lemic2019location}, the generation of the corresponding REMs is ideally to be done with a single tool for the coexisting technologies.

From the above discussion, there is a need for a system for autonomous generation of 3D REMs, which has been recognized in the community.
One of the most promising approaches for such generation is to utilize \acp{UAV} as carriers of the REM-sampling devices~\cite{romero2020aerial,nex2014uav,chakraborty2018skyran}.
In such systems, the samples generated by the REM-sampling devices are correlated with the UAV-originating location at which the measurements are taken, providing the primitive for REM generation.
Such systems have mostly been proposed for outdoor scenarios and \ac{GPS}-enabled large outdoor UAVs. 
These systems are intuitively not suitable for fine-grained generation of REMs indoors because the UAVs are not practically utilizable in smaller and more complex environments.
In addition, the GPS-originating location information is not suitable for the generation of fine-grained REMs due to the relatively large localization errors that such location information unavoidably features~\cite{lymberopoulos2015realistic}. 

To address this issue, we propose a system for generating fine-grained 3D \acp{REM} in small indoor environments.  
The system can be viewed as a toolchain consisting of small-indoor UAVs with fine-grained localization capabilities supported by an \ac{UWB}-based positioning system.
The UAVs are envisioned to serve as the carriers of the REM-sampling receivers and provisioners of location annotation for the sampled measurements.
The location-annotated measurements are envisioned to be streamlined as training inputs to an \ac{ML}-based part of the toolchain, which, once trained, is able to predict the signal quality at locations not visited by the UAVs.    

The most similar existing effort to our work is \cite{automating-wifi-fingerprinting}, where the authors propose a UAV-based system for the generation of a training dataset for 3D Wi-Fi fingerprinting-based indoor localization. 
In contrast to their work focusing solely on Wi-Fi-based fingerprinting, we argue that REMs can be beneficial and utilized more broadly, for example in optimizing the positioning of UAVs serving as mobile relays~\cite{rubin2007placement} or planning the extensions of any wireless networking infrastructure by adding \acfp{AP} or base stations to cover ``dark'' connectivity regions in an environment of interest~\cite{daniel2011using}. 
Hence, our system abides to two additional design requirements compared to \cite{automating-wifi-fingerprinting}. 
The first one is a modular design of the interface between the UAV-based system and a given REM-sampling device. 
This requirement allows for a simple integration of different REM-sampling device (e.g., Wi-Fi, LoRa, BLE, mmWave) with the UAV, extending the REM capabilities beyond the traditional Wi-Fi. 
The second is to guarantee no self-interference between the wireless network used to control the UAVs and the REM-sampling device the UAV is carrying, allowing highly repeatable measurement collections with minimized external influences.
Finally, in our system additional UAVs can be seamlessly integrated into the toolchain, allowing for sequential data collection and scalable REM generation.

We instantiate the proposed system design on two CrazyFlie UAVs to autonomously gather IEEE 802.11b/g/n Wi-Fi beacon frame data in the 2.4 GHz Industrial, Scientific and Medical (ISM) band. 
In a real-world indoor environment (i.e., a living room in a condo apartment), we demonstrate that the UAVs are each able to visit 36 waypoints and collectively gather several thousands of Wi-Fi beacon data samples.
Finally, we instantiate the remaining part of the toolchain on several contemporary ML algorithms, train the system using the data streamlined by the UAVs, and demonstrate a reasonable prediction accuracy at locations not visited by the UAVs.


\section{System Overview}
The REM generation is envisioned to be initiated from the control station by providing a set of waypoints to be visited by a fleet of UAVs operating sequentially. 
The first UAV in the fleet is envisioned to visit a subset of the provided points, with the main limitation on the number of points that can be visited in one run stemming from the constrained battery capabilities of the UAV. 
At each location, the UAV instructs the REM-generating receiver to collect the signal quality indicators of interest, and upon each response it reports the obtained results to the control station. 
This procedure is repeated until all points have been visited by the UAVs or all UAVs in the fleet have depleted their batteries. 
In order to visit the instructed locations, each UAV requires a means for localizing itself, which is supported through its localization system consisting of a client mounted on the UAV and a set of infrastructural devices (i.e., anchors) for the client's localization.  

The main design requirements for the envisioned system include i) accurate location-annotated sampling for streamlined generation of fine-grained 3D REMs, ii) straightforward deployment of the system in unknown complex indoor environments, iii) support for technology-agnostic REM-generating receiver, iv) guaranteed mitigation of self-interference. 

\begin{figure*}[!t]
\begin{minipage}{0.48\linewidth}
\centering  
\includegraphics[width=\linewidth]{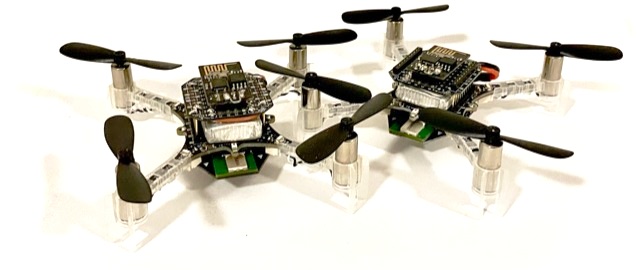}
\caption{Customized Crazyflie 2.1 UAVs}
\label{fig:two_crazyflie_drones}
\end{minipage}%
\hfill
\begin{minipage}{0.48\linewidth}
\centering 
\includegraphics[width=0.98\linewidth]{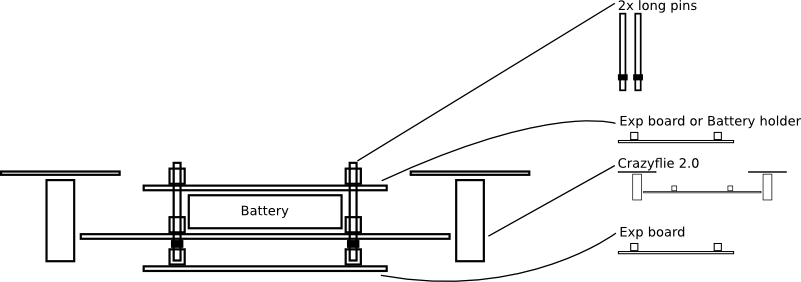}
\caption{Crazyflie expansion boards}
\label{fig:crazyflie_expansion_boards}
\end{minipage}%
\hfill \vspace{-1mm}
\begin{minipage}{0.48\linewidth}
\centering  
\includegraphics[width=\linewidth]{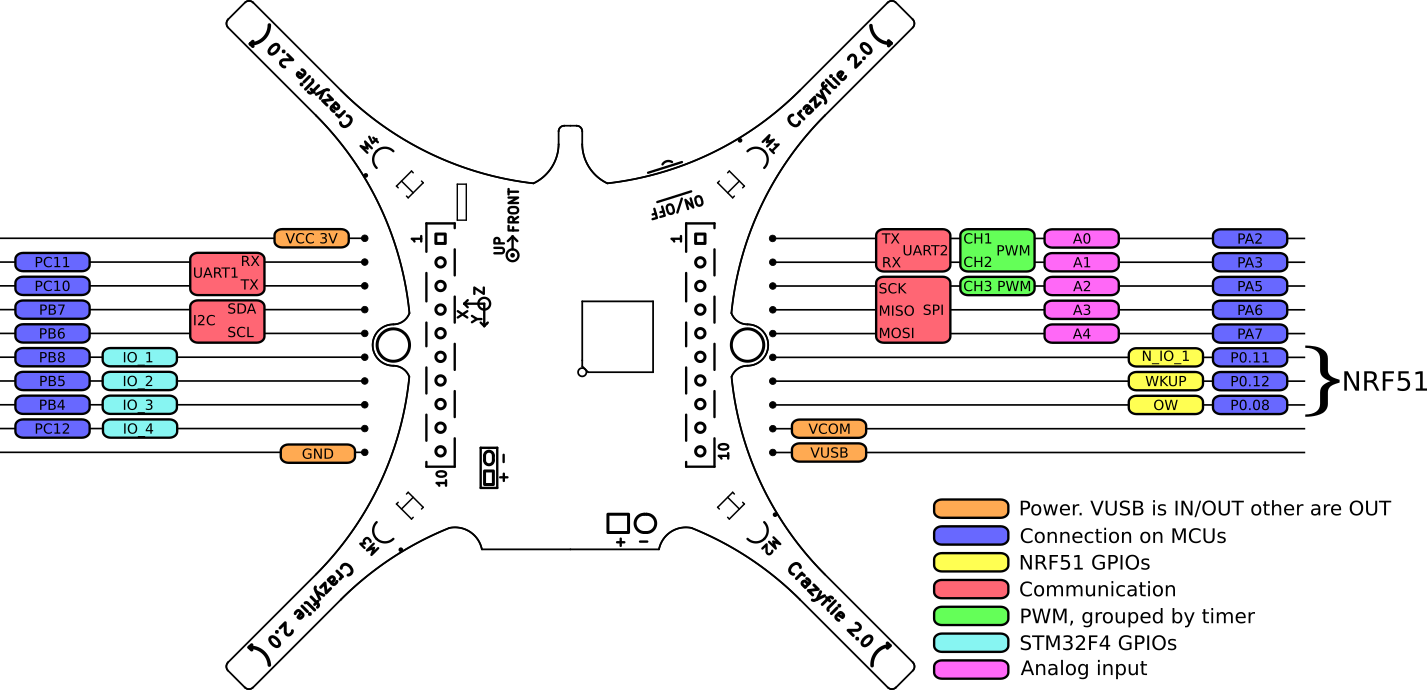}
\caption{Crazyflie interfaces and pin allocation}
\label{fig:crazyflie_pins}
\end{minipage}
\begin{minipage}{0.44\linewidth}
\centering  
\includegraphics[width=\columnwidth]{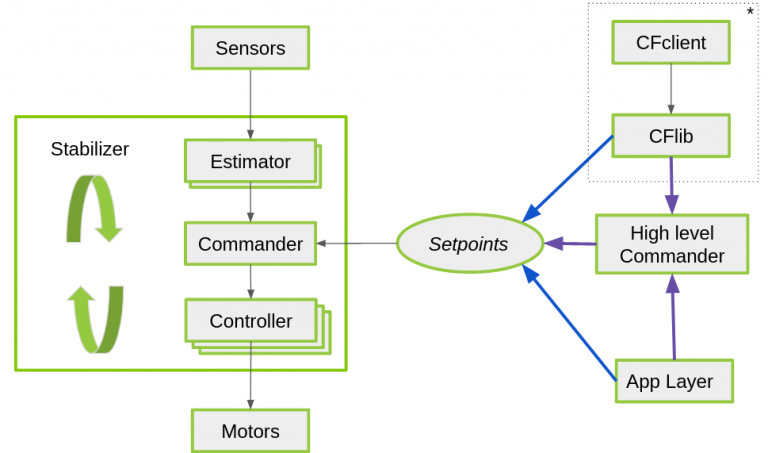}
\caption{The Crazyflie commander framework}
\label{fig:commander_framework}
\end{minipage}%
\vspace{-5mm}
\end{figure*}

We have compared a range of commercial off-the-shelf UAVs based on the design requirements. 
This comparison can be found in the accompanying technical report~\cite{mendes2020report}. 
We have decided to utilize BitCraze Crazyflie 2.1 (Figure~\ref{fig:two_crazyflie_drones}), primarily as it is an open hardware and software platform.
Crazyflie 2.1 UAVs come with a FreeRTOS-based operating system and a radio and Bluetooth LE transceivers for control. 
They also feature an accelerometer, gyroscope, magnetometer, and a high precision pressure sensor through its 10-DOF Inertial Measurement Unit (IMU).
These capabilities can be extended by adding up to two expansion boards or decks as shown in Figure~\ref{fig:crazyflie_expansion_boards}. 
In this work, both expansion slots are used: one for the \ac{LPD}, the other for the integration of an REM-generating receiver.
The Crazyflie provides a set of interfaces that can be addressed over 20 pins for communication with each expansion deck. 
Figure~\ref{fig:crazyflie_pins} documents the pin allocations, where one can choose between an I$^2$C, SPI, STM32, and two UART interfaces. 

\subsection{Interfacing with REM-generating Receivers}

Using the FreeRTOS-flavored Crazyflie 2021.06 firmware release as a basis, a custom driver is responsible for interfacing with an REM-generating receiver. 
The driver should support: i) initializing and ii) checking the state of the REM-generating receiver, iii) instructing the REM-generating receiver to collect a measurement, and iv) enabling parsing of the output of the previous instruction. 
For integration with the UAV, the user is required to provide the driver for the REM-generating receiver to react to the four specified instructions. 
In terms of hardware integration, the user can choose between UART and I2C interfaces available on the Crazyflie UAV.
We argue that such an \textit{integration procedure is straightforward as it is supported by well-known hardware interfaces and a four instructions-long C-flavored driver}, as long as the REM-generating receiver features suitable size (i.e., USB-dongle dimensions) and weigh (up to 20~grams) to be carried by the UAV.

\subsection{UAV Localization}
We use the Crazyflie's \ac{LPS} to provide accurate indoor positioning. This system works with a tag, the \ac{LPD}, and multiple anchors distributed over the volume used for localizing the tag. 
The system can be deployed by simply positioning of the localization anchors, measuring their coordinates relative to a chosen origin, and initializing their automated calibration for synchronizing their transmission schedules.
Once the localization anchors are self-calibrated, they can be used for localizing the UAVs and consequently for the generation of a 3D REM can be initiated.
Given that the procedure for the REM-generating system is relying solely on physical deployment, localization, and initiation of their self-calibration, we argue that \textit{it can be utilized for rapid deployment of our system in complex new environments}. 

The \ac{LPS} is based on the Decawave DWM1000 module and uses \ac{UWB} radio technology for communication and localization.
The tag can estimate its own position based on the UWB signals received from the localization anchors.
The localization is then performed using either the Two-Way Ranging (TWR) procedure or different flavors of the Time Difference of Arrival (TDoA) procedure, the latter featuring slightly better accuracy and supporting simultaneous localization of multiple UAVs.
Regardless of the utilized localization procedure, the \ac{LPS} is able to localize the tag in a 3D environment at the range of about 10~m~\cite{automating-wifi-fingerprinting}.

A  minimum of four localization anchors are required to support UAV localization in 3D environments. 
A Crazyflie equipped with an LPS deck makes use of an extended Kalman filter to estimate its state (orientation and position), its implementation is based on~\cite{extended-kalman}. 
An increase in the number of localization anchors increases the robustness and accuracy of this process.
Hence, Bitcraze advises to use at least six such anchors to mitigate potential negative effects in the deployment environment (e.g., no Line of Sight (LoS) to an anchor, fluctuating tag orientation). 
Chekuri and Won have shown in their tests~\cite{automating-wifi-fingerprinting} that localization with 6 anchors can achieve an accuracy of 9 cm when the UAV is hovering, this is important as we will record the UAV's position and scan while the UAV is holding its position and orientation. 
In summary, \textit{the system is able to generate location-annotated measurements for REM generation with decimeter-level accuracy at 10~m range}.

\subsection{UAV Communication and Control}
\label{communicating_using_the_crazyradio}

Software control of the UAV is done through a Python application that makes use of the Crazyflie Python library. 
This application can communicate with the UAV to send instructions to move to a waypoint, scan for signal quality indicators and other relevant information, parse the results, and store them for later processing.
It runs through the following sequence: i) initialization and instruct the UAV to take-off. For every configured waypoint, make the UAV: ii) move to the waypoint defined with the $\langle x, y, z \rangle$ tuple, iii) initiate an on-demand scan, iv) shutdown the Crazyradio while the scan is running, v) restart the radio connection after the scan is done, vi) fetch the scan results, parse and store them, land the UAV and shut it down, and store the results.
In terms of technology, communicating with and controlling the Crazyflie UAvs remotely can be done using a custom USB dongle called the Crazyradio. 
This radio uses a nRF24LU1 chip providing 126 channels and works with a custom protocol to communicate with the Crazyflie: the Crazyradio RealTime Protocol (CRTP). 
These 126 channels are uniformly distributed over the 2400 MHz - 2525 MHz frequencies.

\begin{figure*}[!t]
\begin{minipage}{0.5\linewidth}
\centering  
\includegraphics[width=\linewidth]{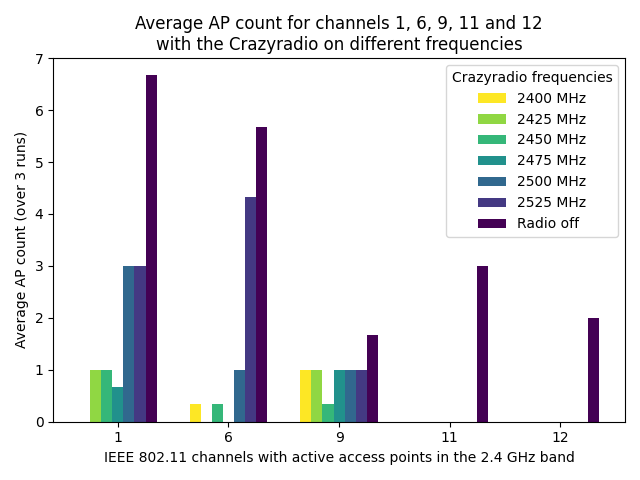}
\caption{The number of APs detected per IEEE 802.11 channel with the Crazyradio set at different frequencies or completely turned off}
\label{fig:crazyradio_interference}
\end{minipage}%
\begin{minipage}{0.46\linewidth}
\centering 
\includegraphics[width=0.93\linewidth]{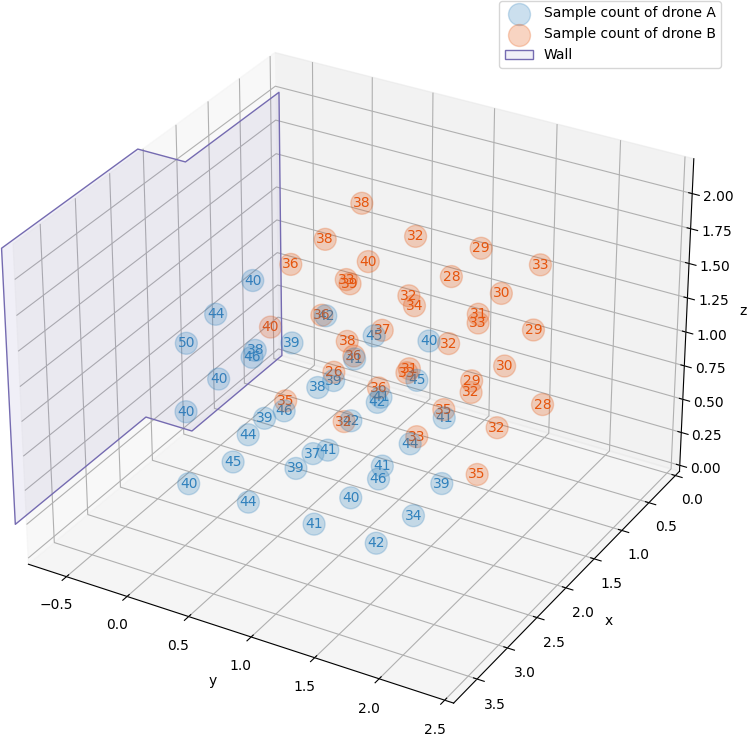}
\caption{Number of samples per UAV and scanned location}
\label{fig:graph_samples_per_scanned_location_per_drone.png}
\end{minipage} \hfil
\begin{minipage}{0.57\linewidth}
\centering 
\includegraphics[width=.96\linewidth]{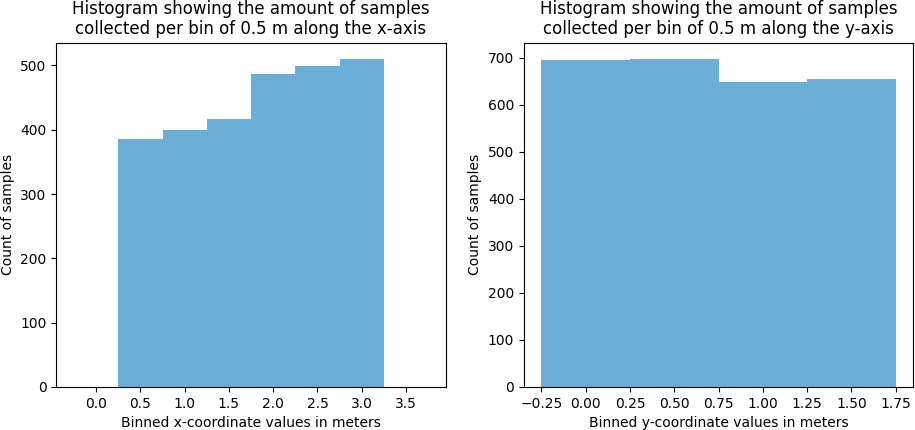}
\caption{Histograms showing the number of samples collected per bin of 0.5 m along the x and y-axis}
\label{fig:graph_histogram_x_samples}
\end{minipage} \hfil
\begin{minipage}{0.40\linewidth}
\centering 
\includegraphics[width=.93\linewidth]{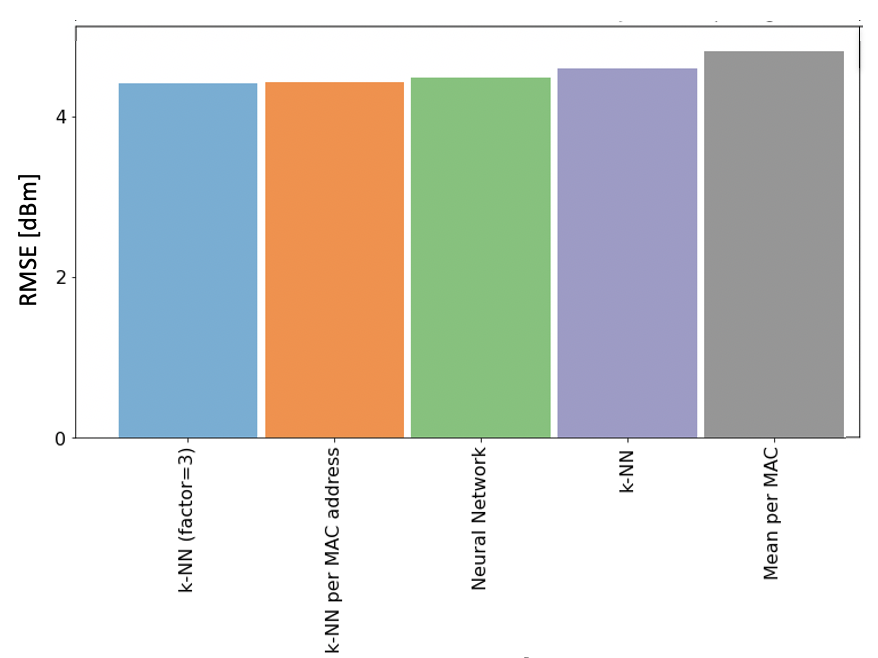}
\caption{RMSE of prediction for different models}
\label{fig:root_mean_square_error_per_regressor}		
\end{minipage}
\vspace{-5mm}
\end{figure*}

There are three potential sources of self-interference from the system toward the REM-generating receiver. 
The first two are the positioning system and the propulsion of UAV's rotational engines, although~\cite{automating-wifi-fingerprinting} showed that these types of interferences have a negligible effect on the REM-generating receiver operating in 2.4~GHz \ac{ISM} band.
The third potential source of interference is the Crazyradio.
The interference generated by the Crazyradio while the \ac{SUT} is scanning for available \acp{AP} at a given location is illustrated in Figure~\ref{fig:crazyradio_interference}. 
The figure shows the average number of APs detected at different 2.4~GHz Wi-Fi channels for 6 operating frequencies of the Crazyradio, as well as in case when the radio was turned off. 
As visible, the interference from the Crazyradio is significant, irrespective of its operating frequency. 

Hence, our experimentation setup features the (default) possibility of automatically turning off the Crazyradio while performing a measurement.
To avoid self-interference, the radio is shut down right before the scan starts and restarted again after the scan has finished. 
The Crazyflie will go into \textit{position hold} mode while the radio is shut down.
Small adjustments were made to the firmware to make working without a radio possible.
First, the \verb|CRTP_TX_QUEUE_SIZE| was increased so that full scan results can be temporarily stored until the radio comes back online and the results can be sent to the controlling application. 
Second, the \verb|COMMANDER_WDT_TIMEOUT_SHUTDOWN| was increased to 10 sec. 
This timeout is a safety measure, if there is no set-point received within this interval, the Crazyflie will shut down under the assumption that something went wrong. 
The default value does not allow to bridge the radio shutdown period.

When the UAV loses its radio connection, it also loses its ability to get new waypoints (i.e., target locations) from the base station. When no new setpoint is received for over 500 ms, the UAV will set its attitude angles (pitch, roll and yaw) to 0 in order to keep itself stabilized. 
Figure~\ref{fig:commander_framework} details how the base station's custom Python client can forward waypoints to the \textit{Commander} in the UAV's firmware by making use of the \textit{CFlib} library.
In order to make the UAV hold its position after shutting down the radio connection, an extra FreeRTOS task was added to the driver of the ESP8266 deck that will feed back the scanning position every 100~ms to the UAV's commander during such a scan. 
This task gets resumed at the start of the scanning task and suspended at the end of it so that it does not interfere with regular waypoint activities. This feedback process results in \textit{the UAV having stability and guaranteed lack of interference while scanning}.


\section{Validation}

\subsection{Collection of 3D REM-generating Measurements}

We demonstrate\footnote{Demonstration video: \url{https://youtu.be/fxDkR-Qat6w}} one instantiation of the proposed system design for small UAVs-supported autonomous generation of fine-grained 3D indoor REMs.
In particular, the system was deployed in a living room of an apartment in a large apartment building in Antwerp, Belgium.
The 3D volume for the UAVs to scan is a rectangular cuboid of 3.74~m long (x-axis), 3.20~m wide (y-axis) and 2.10~m high (z-axis). 
At each of the 8 corners of the cuboid, a localization anchor was placed and manually localized for enabling the UAVs to estimate their locations within the volume.
Once the self-calibration procedure was finished, using the controlling application we have instructed a fleet of Crazyflie UAVs to generate an REM of the volume.

We have opted at generating an REM of 2.4~GHz ISM Wi-Fi. 
This was supported through AI Thinker ESP-01 modules with Espressif Systems ESP8266 Wi-Fi chips.  
The modules were soldered on Crazyflie prototyping decks and enabled to interact with the UAVs as one of their expansion decks. 
Using the Crazyflie 2021.06 firmware release as basis, a custom driver for this ESP-01 module was written. 
This driver communicates with the ESP-01 module over its UART interface by sending AT instructions and parsing the output. 
Since the module is only used to scan for available access points, it suffices that the driver supports just the following AT instructions: i) \verb|AT| - testing AT start-up, ii) \verb|AT+CW_MODE_CUR| - setting the current Wi-Fi mode (to put the module in station mode), iii) \verb|AT+CWLAP| - listing the available APs and scanning for Wi-Fi beacons, \verb|AT+CWLAPOPT| - formatting the output of the \verb|AT+CWLAP| instruction to $\langle ssid, rssi, mac, channel \rangle$ tuples.

Wi-Fi REM was selected due to self-interference with CrazyRadio, as its frequency range overlaps with the 2.4~GHz band that the Wi-Fi modules use. 
In order to get an idea of how pronounced this expected interference is, the Crazyradio was run on different frequencies over its range in 25~MHz increments (2400, 2425, 2450, 2475, 2500 and 2525~MHz). 
At each of these frequencies, 3 access point scans were done using the ESP-01 module on the Crazyflie. To have a baseline for comparison, an additional three scans were done with the Crazyradio turned off. 
The interference generated by the Crazyradio while scanning for access points is illustrated in Figure~\ref{fig:crazyradio_interference}. 
It shows for every Wi-Fi channel the average count of access points that were detected over the 3 runs and shows this for 6 different frequencies of the Crazyradio as well as the radio turned off. 
Wi-Fi channels that did not have any detected \acp{AP} were left out for clarity. 
These scans were done in a short timespan and with the Crazyflie and Crazyradio in a fixed position.
Not only does this data clearly show that the interference of the Crazyradio is significant, it also demonstrates the benefits of our design decision to turn off the Crazyradio while performing REM-generating measurements.

To mitigate interference among UAVs, the UAVs are run in a sequence, not jointly. 
The \ac{LPS} is configured to use the Time Difference of Arrival-based localization procedure.
The Crazyflie is advertised as having a flight time of up to 7~min depending on how it is used. 
This is, however, without the weight and power consumed by the \ac{LPD} and the custom ESP8266 deck. 
There are also several other factors that can influence the UAV's endurance including flight and scan parameters, the choice between TWR and TDoA or the distance to the anchors, to name a few.
To get a notion of the UAV's endurance in a baseline scenario, a UAV was manually flown until it became less responsive and its motions erratic, considering a fully charged standard battery, eight active anchors in TWR mode, periodic scanning mode with an interval of 8~sec, with a beacon scan duration of around 2~sec. The UAV was kept in a steady position about 1~m above ground level for the duration of the test. The UAV was able to perform 36 scans over a timespan of 6~min and 12~sec before it experienced erratic behaviour. 

Obviously, the endurance is expected to be lower when the UAVs visit different locations and scan more frequently.	
With this constraint in mind, 72 locations evenly spread over the volume were identified, with each UAV responsible for scanning 36 of them. 
The UAVs had 4 sec to fly from a location to another and 3 sec for scanning. 
Thus, scanning 36 locations was expected to take at least 4 min and 12 sec. 
Contributing to that the time required for takeoff and landing, the UAVs were expected to operate at their operating limits. 

The UAVs were controlled by a base station, i.e., a laptop running the custom Python client software. 
The client was responsible for sending the UAVs to a waypoint and instructing the scanning. 
Once the scanning at a waypoint was finished, the UAV would send the results back to the client for parsing and storing for further processing. 
The client was configured to be able to control multiple UAVs with a matching set of waypoints and parameters such as radio address, starting position, and yaw. 
While in the paper we show the operation of a two-UAV system, the system can be scaled by simply adding sets of waypoints and above-mentioned parameters.

Using this setup, data was collected for further analysis and processing. A total of 2696 samples were collected, 1495 by UAV A and 1201 by UAV B. During data collection, UAV A was active for 5 min and 3 sec and UAV B for precisely 5 min.
In the collected samples, there were 73 distinct \ac{MAC} addresses and 49 \acp{SSID}, with the mean RSS of around -73~dBm.
When looking at the samples collected per UAV and scanned location (cf., Figure~\ref{fig:graph_samples_per_scanned_location_per_drone.png}), we see no issues with the number of samples collected by UAV B, although this number is generally lower than for UAV A. 
There are environmental factors that can play a role however i) the positive x-axis and negative y-axis point towards the center of the apartment building where we can expect to see more signals, ii) there is a wall segment that is 40~cm wider where UAV B's measurements are taken compared to UAV A, as illustrated in Figure \ref{fig:graph_samples_per_scanned_location_per_drone.png}.

We expected to observe a gradual increase in the number of RSS observations (from more \acp{AP}) toward the center of the building, irrespective of which UAV collected the samples. 
An illustration of this can be seen in Figure~\ref{fig:graph_histogram_x_samples}, which shows a histogram per axis that groups the x and y values in bins of 0.5~m with the height representing the number of samples collected by the UAVs in that bin. 
We can clearly see that the number of samples collected increases with an increasing x-coordinate and a decreasing y-coordinate.

\subsection{Generation of Fine-grained 3D REMs}

A few pre-processing steps were taken before utilizing the data in an \ac{ML} model. 
Since SSIDs can be shared between devices, they were generally not used.
Instead, RSS readings were grouped based on their MAC addresses. 
The timestamps were left out of consideration as well. 
The time difference between the first and last collected sample was less than 10~min, hence considered as irrelevant. 
MAC addresses with less than 16 samples were dropped, since the goal was to predict RSS values of \acp{AP} with sufficient number of measurements. 
Finally, MAC and channel features were considered as categorical and one-hot encoded.
This pre-processing results in 2565 retained samples (131 dropped).

Since the data is locational as it represents signal quality in a 3D space, a k-nearest neighbor regressor was considered. 
As features, the x, y, z coordinates were chosen, as well as the one-hot encoded MAC addresses.
Including the one-hot encoded MAC addresses has the advantage that samples with a different MAC address are considered farther away than similar samples with the same MAC address.
The kNN regressor was configured to use Euclidean distance by setting $metric$=minkowski and $p$=2, as yielded by the the grid search considering an exhaustive set of hyperparameters. 
Similarly, the $weights$ and $n_{neighbors}$ parameters were tuned using a grid search where the optimal values were $weights=distance$ and $n_{neighbors}$=3.

We have considered the samples with different MAC addresses to be further distinguishable by multiplying the one-hot encoded values by the factor of 3 and setting the $n_{neighbors}$ parameter to 16, as yielded by the grid search.
Moreover, as an intuitive alternative to assigning samples with different MAC addresses a greater distance, we considered a kNN estimator per MAC address.
To achieve that, we kept the hyperparameters of these MAC-based regressors the same as previously, and took samples with the same MAC address into account, reducing the feature set to only the x, y, z coordinates. 

The last solution we considered was a neural network. 
In particular, we have built and tested the network for different configurations, including; i) multiple hidden layers with a varying amount of nodes, ii) normalized RSS values, iii) multiple inputs: 1 for the x, y, z coordinates and 1 for the hot-encoded MAC addresses that get combined into a common hidden layer, and iv) different activation functions and optimizers.
The optimized neural network had an input layer for the x, y, z coordinates and the one-hot encoded MAC addresses, sigmoid activation function, hidden layer with 16 fully connected nodes, linear activation function, output layer with a single node for the prediction, and Adam optimizer.

The accuracy of the considered estimators was measured based on the \ac{RMSE} of their predictions.
In order to have an unbiased view on an estimators' predictive capacity, the preprocessed data was split into a training ($75\%$) and test ($25\%$) sets. 
For those estimators that require an additional validation set for tuning their hyperparameters, the validation set was taken out of the training set.
In order to assess more elaborate estimators we used a baseline estimator that always returns the mean per MAC address. 

Figure \ref{fig:root_mean_square_error_per_regressor} shows a comparison of the RMSEs for the different utilized predictors. 
The predictor generally utilizing the mean per MAC address resulted in an RMSE of $4.8107$ dBm. 
The k-NN-based regressors generally yielded slightly better performance than the baseline, with the best performing one being the k-NN algorithm where the one-hot encoded MAC address feature was multiplied by the factor of 3 and $n_{neighbors}$=16, resulting in the RMSE of $4.4186$ dBm.
Finally, the optimally tuned neural network with a single hidden layer of 16 nodes yielded the RMSE of 4.4870 dBm. 
While this is slightly better compared to the baseline, it does fall short of the best k-nearest neighbors solution, as indicated in the figure.
This collection of regressors was used with a relatively small set of collected samples, which explains their comparable performance. Nonetheless, we believe that this exercise demonstrates the toolchain-like utilization of the proposed system with different ML tools.


\section{Conclusion}

We have proposed a system for autonomous collection of \acf{3D} \acfp{REM}-generating measurements. 
The main components of the system include UWB localization-enabled UAVs acting as the carriers of REM-generating devices. 
The location-annotated measurements for 3D REM generation are then streamlined in a \acf{ML} entity for fine-grained prediction of signal quality indicators at locations not visited by the UAVs.

Future work will include instantiating the system on a larger number of UAVs, followed by exhaustive data collection for a set of representative environments and different wireless technologies.
These datasets will be used for deriving the fundamental limitations on the density of 3D REMs.
The UAV localization is currently utilizing UWB, hence the REM-generating device has to operate in a different frequency band to mitigate the self-interference effects.
This represents an obvious limitation of our system. To address this issue, future work will focus on integrating the BitCraze's infrared system called \textit{Lighthouse} for UAV localization, which features comparable precision, while requiring less anchors and being cheaper. In addition to further self-interference mitigation, this effort is expected to make the system even easier to deploy.

\vspace{-0.5mm}
\section*{Acknowledgments}
Filip Lemic was supported by the EU Marie Curie project "Scalable Localization-enabled In-body Terahertz Nanonetwork" (nr. 893760). In addition, this work received support from the University of Antwerp’s Research Fund (BOF).

\renewcommand{\bibfont}{\footnotesize}
\printbibliography

\end{document}